\documentclass[12pt]{iopart}
\usepackage{epsfig}

\usepackage{amssymb,amsmath}
\usepackage{graphics}
\usepackage{dcolumn}
\usepackage{bm}
\usepackage{epsfig}
\usepackage{array}

\newcommand{\GF}{{\rm Green's function\ }}

\newcommand{\NE}{{\rm non-equilibrium\ }}

\begin{document}
\jl{3}

\title[Dyson equations for time-ordered Green's functions]
{Dynamical equations for time-ordered Green's functions: from the
Keldysh time-loop contour to equilibrium at finite and zero temperature}

\author{H. Ness, L. K. Dash}
\address{Department of Physics, University of York, Heslington, York YO10 5DD, UK}
\address{European Theoretical Spectroscopy Facility (ETSF)}

\begin{abstract}
We study the dynamical equation of the time-ordered
Green's function at finite temperature. 
 We show that the time-ordered Green's function obeys a conventional Dyson equation
  only at equilibrium and in the limit of zero-temperature. 
In all other cases,
  i.e. finite-temperature at equilibrium or non-equilibrium, the time-ordered 
  Green's function obeys instead a modified Dyson equation. The derivation of this result
  is obtained from the general formalism of the non-equilibrium Green's functions 
  on the Keldysh time-loop contour.  At equilibrium, our result is fully consistent
with the Matsubara temperature Green's function formalism and also justifies rigorously
the correction terms introduced in an {\it ad hoc} way with Hedin and Lundqvist.
Our results show that one should use the appropriate dynamical equation for the
time-ordered Green's function when working beyond the equilibrium zero-temperature limit.
\end{abstract}
 

\maketitle

\section{Context}
\label{sec:intro}

Equilibrium, zero-temperature and finite-temperature Green's functions (GF) techniques
based on many-body perturbation theory are widely used in
electronic-structure and total-energy calculations.
They are central for the calculation of the thermodynamical properties of many-body systems, 
as well as for the calculation of linear responses of systems under small time-dependent
(or not) perturbations 
\cite{Pines:1961,Kadanoff:1962,Abrikosov:1963,Fetter:1971,Rammer:1986,Mahan:1990,Bruus:2004,Dickhoff:2008}.

Conventional equilibrium many-body GF approaches, at zero temperature, concentrate on the time-ordered GF
and determine its dynamical equation (in the form of an equation of motion, or its integral
form known as the Dyson equation) using a self-energy which includes all the many-body interaction
effects. The Matsubara formalism using time-ordered GF with imaginary time arguments (or imaginary
Matsubara frequency in the dual Fourier representation) has been developed to deal with the
finite temperature condition. 
Interestingly, not much has been done at finite-temperature for the time-ordered GF in real time
or real frequency representation.

However, when one wants to study systems driven out of equilibrium by an 
external ``force'', as for example in the context of electron or heat current 
flow, or for any system driven by an external electromagnetic field
(whether time-dependent or not), one needs to extend the equations for the
dynamics of the quantum many-body interacting system to the
non-equilibrium conditions.

For this, the non-equilibrium Green's function (NEGF)
technique\cite{Kadanoff:1962,Keldysh:1965,Danielewicz:1984,Chou:1985,Wagner:1991,vanLeeuwen:2006,Rammer:2007}
has been widely used to calculate non-linear charge and heat transport properties of
solids, mesoscopic and nanoscopic systems.
Also known as the closed time-path formalism
\cite{Schwinger:1961,Chou:1985}, the NEGF formalism relies on an ``artificial'' time
parameter that runs on a mathematically convenient time-loop contour
(plus eventually an imaginary time for taking into account the initial
correlation and statistical boundary conditions).  It is a formal
procedure that only has a direct physical meaning when one projects
back the time parameters of the time-loop contour onto real times.  It
was introduced because it allows one to obtain self-consistent
Dyson-like equations for the Keldysh Green's function using
Schwinger's functional derivative technique.  Transforming the Dyson
equation to real time by varying the Keldysh time parameter over the
time-loop contour results in a set of self-consistent equations
for the different non-equilibrium Green's functions (advanced/retarded
or lesser/greater).
The NEGF technique is general and can treat non-equilibrium as well as
equilibrium conditions, at zero and finite temperatures, within
a single framework.

In this paper we focus on the dynamical equation of the time-ordered
GF with real time argument, and we show that the time-ordered GF obeys
a conventional Dyson equation only in a limiting case: at equilibrium and zero
temperature. In any other cases, i.e.\ equilibrium at finite
temperature or non-equilibrium, the time-ordered GF obeys instead a \emph{modified}
Dyson equation containing correction terms.
These correction terms were introduced in an {\it ad hoc} way in
the seminal work of Hedin and Lundqvist \cite{Hedin:1969}. 

We derive the modified Dyson equation for the time-ordered GF
in a natural way from the Keldysh time-loop contour approach. We also show that
for the equilibrium case at finite temperature, such correction terms
can be obtained by a proper treatment of the imaginary time Matsubara
finite-temperature Green's functions.

The paper is organized as follows: in section \ref{sec:NEGF-EOM}, we derive
the modified Dyson equation for the time-ordered GF from the time-loop contour
formalism. We then concentrate on the equilibrium condition. 
Section \ref{sec:GFatT_HandL} shows explicitly the relation between Hedin
and Lundqvist's seminal work on real-time GFs at finite temperature and our
result. 
In section \ref{sec:MatsubaraGF}, we show that our result can also be obtained
in by using Matsubara temperature GFs and a proper treatment of the analytical
continuation rules.
We conclude and discuss the importance of our result in section \ref{sec:ccl}.
\ref{app:analyticalcontinuation} and \ref{app:GFSEdefinitions}
contain important and useful relations for the GFs and self-energies on 
the time-loop contour. In \ref{app:Hubbardmodel} we provide a concrete
example of our general results for a specific system: the single site Hubbard 
model at finite temperature.

\section{Non-equilibrium Green's functions on the Keldysh contour $C_K$}
\label{sec:NEGF-EOM}

\begin{figure}
\vspace{1cm}
\begin{center}
\setlength{\unitlength}{5cm}
\begin{picture}(1.5,0.1)
\thicklines
\put(1.15,0.12){\oval(0.1,0.12)[r]}
\put(0.25,0.18){\line(1,0){0.9}}
\put(0.01,0.18){\footnotesize{$-\infty_+$}}
\put(1.15,0.20){\footnotesize{\rm{branch $(+)$}}}
\put(1.15,0.06){\vector(-1,0){0.9}}
\put(0.01,0.06){\footnotesize{$-\infty_-$}}
\put(1.25,0.07){\footnotesize{\rm{branch $(-)$}}}
\put(0.2,0.01){\vector(1,0){1.2}}
\put(0.4,0){$\vert$}
\put(0.4,-0.1){\footnotesize{$t_1$}}
\put(0.8,0){$\vert$}
\put(0.8,-0.1){\footnotesize{$t_2$}}
\put(1.2,0){$\vert$}
\put(1.2,-0.1){\footnotesize{$t$}}
\put(0.38,0.165){$\times$}
\put(0.38,0.22){$\tau_1$}
\put(0.78,0.045){$\times$}
\put(0.78,0.10){$\tau_1'$}
\put(1.42,0){\footnotesize{\rm{time}}}
\end{picture}
\end{center}

\caption{The time-loop Keldysh contour $C_K$ for an ``artificial'' time parameter ordering on a 
forward time evolution ($+$) branch followed by a backward time evolution ($-$) branch, the ``turning'' point
$t$ is taken as $t\rightarrow \infty$.}
\label{fig:CK}
\end{figure}
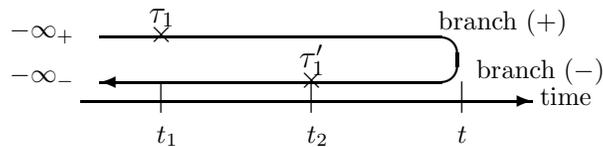

The Green's function on the Keldysh contour $C_K$ is defined as 
\begin{equation}
\label{eq:defGFonCK}
G(1,2)	= - {\rm i} \langle \mathcal{T}_{C_K} \Psi(1) \Psi^\dag(2) \rangle,
\end{equation}
where $(1,2)$ stands for a composite index for space
$\mathbf{x}_{1,2}$ and time $\tau_{1,2}$ on the time-loop
contour $C_K$. 
The time-loop contour $C_K$ consists of a chronological time evolution ($+$) branch 
followed by an anti-chronological time evolution ($-$) branch \cite{Keldysh:1965,Danielewicz:1984,Chou:1985,vanLeeuwen:2006},
as shown in Figure \ref{fig:CK}.
The time ordering $\mathcal{T}_{C_K}$ of the product
of fermion creation ($\Psi^\dag$) and annihilation ($\Psi$) quantum fields
is performed on the time-loop contour $C_K$.

The Green's function obeys the equation of motion on the contour $C_K$ \cite{Danielewicz:1984,vanLeeuwen:2006}:
\begin{equation}
\label{eq:EOM}
[i\partial_{\tau_1} - \bar{h}_0(1)]G(1,1^\prime) =  \delta(1 - 1^\prime) +
\int {\rm d}3 \Sigma(1,3)G(3,1^\prime),
\end{equation}
and the adjoint equation reads
\begin{equation}
\label{eq:EOMadj}
[-i\partial_{\tau_1^\prime} - \bar{h}_0(1^\prime)]G(1,1^\prime) =
\delta(1 - 1^\prime) + \int {\rm d}2 G(1,2) \Sigma(2,1^\prime),
\end{equation}
where $\bar{h}_0(1)$ is the non-interacting Hamiltonian.

Hence, in a integral form, $G(1,1^\prime)$ follows a Dyson-like equation
\begin{equation}
\label{eq:DysonGCK}
G(1,2) = g(1,2) + \int {\rm d}3 {\rm d}4\ g(1,3) \Sigma(3,4) G(4,2) ,
\end{equation}
where $g(1,2)$ is the \GF for the non-interacting system, i.e.
$[i\partial_{\tau_1} - \bar{h}_0(1)]g(1,2) =  \delta(1 - 2)$,
and the time integrations run on $C_K$.
Eq.~(\ref{eq:DysonGCK}) is valid provided that, in the remote past,
the system is unperturbed and non-interacting and that the adiabatic assumption 
for switching on the interaction is fulfilled, as usually assumed within
the Keldysh formalism \cite{Keldysh:1965}.
Furthermore the non-interacting \GF $g(1,2)$ satisfies the
Kubo-Martin-Schwinger boundary conditions (for example see Eq.~(3.16) in 
\cite{vanLeeuwen:2006}).

Upon the position of the time arguments $(\tau_1,\tau_2)$ on $C_K$, one obtains
four different projections onto the real (normal) time arguments $(t_1,t_2)$, i.e.
the four ``Keldysh components'' of $G(1,1^\prime)$ : 
$G^{\eta_1 \eta_2}(\mathbf{x}_{1}t_1,\mathbf{x}_{2}t_2)$
with $t_i$ on the $\eta_i=\pm$ branch of $C_K$.

So $G^{++}(t_1,t_2)= - {\rm i} \langle \mathcal{T} \Psi(1) \Psi^\dag(2) \rangle$ is
the conventional time-ordered GF (though in the \NE conditions), and 
$G^{-+}(t_1,t_2)= - {\rm i} \langle \Psi(1) \Psi^\dag(2) \rangle = G^>(1,2)$
and 
$G^{+-}(t_1,t_2)= {\rm i} \langle \Psi^\dag(2) \Psi(1) \rangle = G^<(1,2)$
are automatically ordered knowing that any time on the branch $(-)$ is later than any time
on the branch $(+)$ on $C_K$.
Note that the equation of motion for $G^\lessgtr(1,2)$ does not contain the source term
$\delta(1 - 2)$ since the time $\tau_1$ and $\tau_2$ are on different branches, and in
the non-interacting case, one has $[i\partial_{\tau_1} - \bar{h}_0(1)]g^\lessgtr(1,2) =0$,
and $[-i\partial_{\tau_2} - \bar{h}_0(2)]g^\lessgtr(1,2) =0$.

Using the relationships between the different GFs given in Appendix \ref{app:GFSEdefinitions},
one can calculate the integral equations for the different Green's functions
\cite{Kadanoff:1962,Keldysh:1965,Danielewicz:1984,Chou:1985,Wagner:1991,vanLeeuwen:2006,Rammer:2007}, 
and one finds that the retarded and advanced $G^{r,a}$ Green's functions obey a Dyson-like equation:
\begin{equation}
\label{eq:DysonGra}
G^{r}(1,2) = g^{r}(1,2) + \int {\rm d}(34)\ g^{r}(1,3) \Sigma^{r}(3,4) G^{r}(4,2) ,
\end{equation}
(likewise for $G^a$)
where the time integrations are on the real times $\int_{-\infty}^{+\infty} {\rm d}t_i$.

The lesser and greater $G^\lessgtr$ GFs obey the following quantum kinetic equation:
\begin{equation}
\label{eq:DysonGlessgtr}
G^\lessgtr(1,2) = \int {\rm d}3 {\rm d}4 \
(1+G^r \Sigma^r) g^\lessgtr (1+\Sigma^a G^a) + G^r \Sigma^\lessgtr G^a \ ,
\end{equation}
given in a compact notation. Eq.(\ref{eq:DysonGlessgtr}) is valid within the conventional
assumption of the Keldysh time-loop contour formalism where the interactions are switched
on adiabatically. The most general solution for the lesser/greater \GF can be found in 
Appendix A of Ref.~\cite{Stefanucci:2004a}.

\subsection{The time-ordered Green's function on $C_K$}
\label{sec:Gt_on_CK}

Following the rules of decomposition of the time-loop contour $C_K$ onto real-time
given in Appendix \ref{app:analyticalcontinuation}, one can expand the
equation for the time-ordered GF $G^t= G^{++}$ as follows (using compact notation)
\begin{equation}
\label{eq:app_G++}
\begin{split}
G^{++} & = g^{++} + (g \Sigma G)^{++} \\
       & = g^{++} 
	 + g^{++} \Sigma^{++} G^{++} \\
       & - g^{++} \Sigma^< G^>
	 - g^< \Sigma^> G^{++}
	 + g^< \Sigma^{--} G^{++} ,
\end{split}
\end{equation}
hence we get
\begin{equation}
\label{eq:app_G++bis}
\begin{split}
G^t 	= & g^t \left( 1+  \Sigma^t G^t - \Sigma^< G^> \right) \\
& - g^< \left( \Sigma^{\tilde t} G^> - \Sigma^> G^t \right) \ .
\end{split}
\end{equation}
After further manipulation, using \ref{app:analyticalcontinuation} and \ref{app:GFSEdefinitions},
we can show that the second line in Eq.~(\ref{eq:app_G++bis}) becomes
$g^< \left( \Sigma^{\tilde t} G^> - \Sigma^> G^t \right) = g^< (\Sigma G)^>$. Hence the full
equation for the time-ordered GF is
\begin{equation}
\label{eq:app_Gt_correct}
G^t = g^t \left( 1+ \Sigma^t G^t - \Sigma^< G^> \right) +  g^< (\Sigma G)^> \ .
\end{equation}

From the equation of motion of GF on the Keldysh contour, we find
\begin{equation}
\label{eq:extra}
g^< (\Sigma G)^> \equiv \int {\rm d}2\ g^<(1,2) [{\rm i}\partial_{t_2} - \bar{h}_0(2)] G^>(2,3) \ ,
\end{equation}
and by using the (adjoint) equation of motion of the non-interacting GF $g^<$, i.e.
$[-{\rm i}\partial_{t_2} - \bar{h}_0(2)] g^>(1,2)=0$,
we can integrate Eq.~(\ref{eq:extra}) by part and find that
\begin{equation}
\label{eq:extra_2}
[g^< (\Sigma G)^>](1,3) \equiv \int {\rm d}x_2 \
{\rm i} [g^<(1,x_2 t) G^>(x_2 t ,3)]_{t=+\infty} - {\rm i} [g^<(1,x_2 t) G^>(x_2 t ,3)]_{t=-\infty} \ ,
\end{equation}
which represents a boundary condition term for the time variable (after all we are solving 
a first order differential equation in time).

Hence Eq.~(\ref{eq:app_G++bis}) is transformed into
\begin{equation}
\label{eq:app_Gt_last_bis}
\begin{split}
G^t(1,2) = g^t(1,2) + \int {\rm d}(34) \ g^t(1,3) \left( \Sigma^t(3,4) G^t(4,2) - \Sigma^<(3,4) G^>(4,2) \right) \\
- \int {\rm d}x_2 \
{\rm i} [g^<(1,x_2 t) G^>(x_2 t ,3)]_{t=+\infty} - {\rm i} [g^<(1,x_2 t) G^>(x_2 t ,3)]_{t=-\infty},
\end{split}
\end{equation}
or equivalently (in compact notation)
\begin{equation}
\label{eq:app_Gt_last_ter}
G^t = g^t \left( 1+ \Sigma^t G^t - \Sigma^< G^> \right) +  \ {\rm boundary\ condition}.
\end{equation}

Eq.~(\ref{eq:app_Gt_last_bis}), equivalently Eq.~(\ref{eq:app_G++bis}), 
is the most general result for the
time-ordered Green's function and is valid for the non-equilibrium
case at finite and zero temperature. It also clearly shows that the 
time-ordered Green's function $G^t(1,2)$ does not follow a Dyson-like 
equation. However this is not a closed equation, in contradistinction
with the Dyson equations for the advanced and retarded components, since
it also involves greater and lesser components for the GF and the self-energy.
In the general non-equilibrium conditions, one rather works 
with the closed equations obtained for the retarded and advanced GF 
which are related to spectroscopic properties, and with the quantum kinetic
equations obtained for the lesser and greater GF.
Though at equilibrium, the lesser and greater components of GF or a self-energy
are related to the respective spectral functions and the equilibrium distribution.
Hence we recover closed equations for the time-ordered GF. 

We also establish in Section~\ref{sec:MatsubaraGF}
the formal links with the approach of Hedin and Lundqvist and with the 
Matsubara temperature GF formalism.

It would be interesting to find some conditions under which the the boundary condition 
term in Eq.~(\ref{eq:app_Gt_last_ter}) vanishes. 
At equilibrium, because of the principle of adiabatic
switching (on/off) of the interactions in the remote past/future, the initial state
returns (up to a phase factor) to itself in the far future. Therefore it would be
tempting to consider that the the two terms
defining the boundary condition are the same at $t=\pm \infty$ and that the boundary
condition term vanishes. However this is not that case, as shown in Section~\ref{sec:MatsubaraGF}.

At non-equilibrium, the problem is different and the
solution may depend on the size of the system. 
For an infinite system (a large system in the thermodynamic limit), in some cases,
we would like to use the so-called memory-loss theorem 
\cite{Stefanucci:2004a,Doyon:2006,Stefanucci:2007}: the long time
limit of the lesser (or greater) components of the GF or of the self-energy vanish.
Hence in that case, the boundary condition term may vanish (though this needs
to be proven).
This corresponds to cases when the system ``thermalise with the environment''.
For finite s‪ize systems at 
non-equilibrium, no ``thermalisation'' is possible and the boundary condition term play an 
important role in the solution of dynamical equation of the time-ordered GF.

Before turning onto the equilibrium condition, let us make two further general
remarks.
One could note that from the definition $G^r(1,2)=G^t(1,2)-G^<(1,2)$ 
(see Appendix \ref{app:GFSEdefinitions})
or equivalently $G^t(1,2)=G^r(1,2)+G^<(1,2)$, one could have gueseds that $G^t$ does
not obey a Dyson equation since $G^r$ does, but not $G^<$
\cite{footnote2}.
 Futhermore we have also checked that we can also recover, as expected, 
our result Eq.~(\ref{eq:app_Gt_last_ter}) by using the Kadanoff-Baym formalism which
is somewhat different but equivalent to the Keldysh time-loop approach 
\cite{footnote3}.

\subsection{At equilibrium}
\label{sec:equi}

We now turn onto the special case of equilibrium for which the Green's functions techniques
based on many-body perturbation theory were orginally derived.

At equilibriumn, as in the steady state, all quantities $X(t,t')$ (GFs and self-energies)
depend only on the time difference $X(t-t')$. One can then Fourier transform the quantity
$X(t-t')$ and work in the real-frequency representation $X(\omega)$.

The main result of the paper, Eq.~(\ref{eq:app_Gt_last_bis}), is recast as:
\begin{equation}
\label{eq:app_Gt_w}
\begin{split}
G^t(\omega) = g^t(\omega) \left( 1+ \Sigma^t(\omega) G^t(\omega) - \Sigma^<(\omega) G^>(\omega) \right) 
- g^<(\omega) (\Sigma(\omega) G(\omega))^> ,
\end{split}
\end{equation}
in compact notation (i.e. omitting the explicit integrals over space).

It should be noticed that the last term $\Sigma^< G^>$ in Eq.~(\ref{eq:app_Gt_last_bis})
satisfies the detailed balance
equation at equilibrium \cite{Danielewicz:1984}: 
$\Sigma^<(\omega) G^>(\omega) = \Sigma^>(\omega) G^<(\omega)$.  

Eq.~(\ref{eq:app_Gt_last_bis}) can also be recast as
\begin{equation}
\label{eq:app_Gt_w_Hedinform}
\left[ [g^t(\omega)]^{-1} - \Sigma^t(\omega) \right] G^t(\omega) + \Sigma^<(\omega) G^>(\omega) = 1 \ ,
\end{equation}
ignoring the boundary conditions term.

Furthermore, at equilibrium, the lesser and
greater components of either a GF or a self-energy ($X^\lessgtr$)
can be expressed in terms of the corresponding advanced and retarded
quantity and a distribution 
function 
\cite{Abrikosov:1963,Fetter:1971,Lipavski:1986,Rammer:1986,Meden:1995,Bruus:2004,Vasko:2005,Ness:2010,Kita:2010},
i.e.
\begin{equation}
\label{eq:NEdistrib}
X^\lessgtr(\omega) =
-f^{\lessgtr,{\rm eq}}(\omega) (X^r(\omega)-X^a(\omega)) ,
\end{equation}
and spectral function $A^X=(X^r(\omega)-X^a(\omega))/(-2\pi i)$.

At equilibrium and for a
system of fermions, $f^{<,{\rm eq}}$ is given by the Fermi-Dirac distribution
function $f^{\rm eq}(\omega)=[1+\exp\beta(\omega-\mu^{\rm eq})]^{-1}$
and $f^{>,{\rm eq}}=f^{\rm eq}-1$ (with $\beta=1/kT$). Eq.~(\ref{eq:NEdistrib})
follows from the Kubo-Martin-Schwinger relationship between the lesser and greater
component at equilibrium: 
$X^{>,{\rm eq}}(\omega)\ =\ - {\rm e}^{\beta(\omega-\mu^{\rm eq})} X^{<,{\rm eq}}(\omega)$.

Hence we can easily see that 
\begin{equation}
\label{eq:SigmaLessGgtr}
\Sigma^<(\omega) G^>(\omega) = (2\pi)^2 f^{\rm eq}(\omega) (1-f^{\rm eq}(\omega) ) A^\Sigma(\omega) A^G(\omega) .
\end{equation}

Therefore in the limit of zero temperature, there is no contribution from $\Sigma^< G^>$ since
$f^{\rm eq} (1-f^{\rm eq})=0$ (this property holds for any product $X^< Y^>$, and hence for
the boundary condition term $g^< (\Sigma G)^>$ ), and the time-ordered GF $G^t$ obeys a 
conventional Dyson-like equation
$G^t(\omega) = g^t(\omega) [ 1+ \Sigma^t(\omega) G^t(\omega) ]$. 

It is only in the case of equilibrium
at zero temperature that $G^t$ follows a Dyson equation otherwise at finite temperature $G^t$ obeys
a modified Dyson equation, Eq.~(\ref{eq:app_Gt_w_Hedinform}), including the extra term $\Sigma^< G^>$
and the boundary condition term $g^< (\Sigma G)^>$.

In the following section, we are going to show how Hedin and Lundqvist noticed the necessity to introduce
correction terms in the real-time Green's function at finite temperature \cite{Hedin:1969}. 
They found it necessary to introduce extra terms in the Dyson equation of the 
real-time Green's function at finite temperature to recover the expected behaviour of the 
independent particle case. The terms introduced {\it ad hoc} by Hedin and Lundqvist correspond
exactly to the $\Sigma^< G^>$ term that we have obtained in a natural way from the use
of the time-loop contour.

\section{Real-time Green functions at finite temperature}
\label{sec:GFatT_HandL}

In this section, we retrace the analysis made by Hedin and Lundqvist (section 17 of Ref.~\cite{Hedin:1969})
on the temperature Green's functions.
The definition for the real-time GF at finite temperature follows from the zero temperature case
by using an ensemble average for the time-ordered product of fermion operators instead of a ground
state expectation value, i.e.
\begin{equation}
\label{eq:Gt_T}
G(1,2)= - {\rm i} {\rm Tr} \left[\rho_G \mathcal{T} \Psi(1) \Psi^\dag(2)\right] ,
\end{equation}
where the statistical operator (density matrix) $\rho_G = Z_G^{-1} \exp-\beta(H-\mu^{\rm eq} N)$
and the partition function $Z_G = {\rm Tr} [\rho_G]$ are given for the grand canonical ensemble.

After Fourier transform, we can write the time-ordered GF using a spectral
representation just as in the zero temperature case \cite{Abrikosov:1963,Fetter:1971,Hedin:1969}:
\begin{equation}
\label{eq:gt_T_w}
G(\omega) = \int {\rm d}\omega' A(\omega') 
 \left[ \frac{1-f^{\rm eq}(\omega')}{\omega-\omega'+{\rm i}\eta} 
      + \frac{f^{\rm eq}(\omega')}{\omega-\omega'-{\rm i}\eta}
 \right] ,
\end{equation}
where we omit the spatial variables.

From Eq.~(\ref{eq:gt_T_w}), we can obtain the real and imaginary parts of $G(\omega)$:
\begin{equation}
\label{eq:Regt_T_w}
\Re e G(\omega) = PV \int {\rm d}\omega' \frac{A(\omega') }{\omega-\omega'} ,
\end{equation}
and
\begin{equation}
\label{eq:Imgt_T_w}
\Im m G(\omega) = \pi [2 f^{\rm eq}(\omega)-1] A(\omega) \ .
\end{equation}

Hedin and Lundqvist noticed that by considering the conventional Dyson equation for the
time-ordered GF: $[\omega - \epsilon - \Sigma(\omega)] G(\omega) = 1$, one
derives the following spectral function
\begin{equation}
\label{eq:wrongAw}
A(\omega) = \frac{B(\omega)}{[\omega - \epsilon - \Re e \Sigma(\omega)]^2+\pi^2 B(\omega)^2 [2 f^{\rm eq}-1]^2 } ,
\end{equation}
with the spectral representation for the self-energy
$\Im m \Sigma(\omega) = \pi [2 f^{\rm eq}(\omega)-1] B(\omega)$.

However Eq.~(\ref{eq:wrongAw}) does not have the proper limit in $\delta(\omega - \epsilon)$ for the independent
particle case where $\Re e \Sigma(\omega)=0$ and $B(\omega) \rightarrow 0$.

To palliate this discrepancy, Hedin and Lundqvist introduced {\it ad hoc} a correction term in the conventional
Dyson equation that $G$ and $\Sigma$ must obey. Their modified Dyson equation reads
\begin{equation}
\label{eq:GtDyson_fromHedin}
[\omega - \epsilon - \Sigma(\omega)] G(\omega) +(2\pi)^2 f^{\rm eq} (1-f^{\rm eq} ) A(\omega) B(\omega) =1 .
\end{equation}
From this modified Dyson equation, one can calculate the imaginary part of $G(\omega)$ and
obtain the spectral function $A(\omega)$
\begin{equation}
\label{eq:Aw}
A(\omega) = \frac{B(\omega)}{[\omega - \epsilon - \Re e \Sigma(\omega)]^2+\pi^2 B(\omega)^2 } ,
\end{equation}
as for the zero temperature limit, it has the proper limit $A(\omega)\rightarrow \delta(\omega - \epsilon)$ for 
the limit $\Re e \Sigma(\omega)=0$ and $B(\omega) \rightarrow 0$.

Hedin and Lundqvist conclude ``{\it that the rather messy definition of $\Sigma$ in Eq.~(\ref{eq:GtDyson_fromHedin})
shows that it is better if, at finite temperature, we can work with concepts other that the straight-forward
generalisation in Eq.~(\ref{eq:Gt_T}) of the zero temperature Green's function. One possibility is to use the 
Matsubara technique with temperature Green's functions, or the close related Martin and Schwinger technique ...}''
\cite{Hedin:1969}.

It is interesting to note that the ``messy'' definition Eq.~(\ref{eq:GtDyson_fromHedin}) is actually the
correct result for the time-ordered GF $G^t$ at finite temperature. Eq.~(\ref{eq:GtDyson_fromHedin}) 
is strictly equivalent to Eq.~(\ref{eq:app_Gt_w_Hedinform}) when one notices that in the definition of
$\Sigma^< G^>$ in Eq.~(\ref{eq:SigmaLessGgtr}) one deals with the same spectral functions, i.e.
$A\equiv A^G$ and $B\equiv A^\Sigma$.

So by using the Keldysh time-loop contour, we have found in a natural way, almost straightforward, the proper
modified Dyson equation for $G^t$ at equilibrium and this, without the need of introducing correction terms 
with no proper justification, other than the need to recover the correct behaviour for  $A^G(\omega)$ in the
non-interacting case.

To conclude this section, one could rephrase our result by saying that the time-ordered GF $G^t$
at equilibrium and finite temperature obeys a Dyson equation $G^t = g^t + g^t \bar\Sigma G^t$ only 
if one defines the corresponding self-energy $\bar\Sigma$
as $\bar\Sigma = \Sigma^t - \Sigma^< G^> (G^t)^{-1}$.

\section{Recovering $G^t$ from the equilibrium formalism and imaginary-time temperature Green's functions}
\label{sec:MatsubaraGF}

In the next section, we show that the proper modified Dyson equation for $G^t$ can be
obtained from the knowledge of the spectral function $A^X(\omega) = (X^r(\omega)-X^a(\omega))/-{\rm i}2\pi$,
with $X$ being a GF or a self-energy.
We also discuss the connection of our result with the Matsubara temperature Green's functions.

We extract the spectral function $A^G$ using the Dyson equations for $G^{r,a}$:
\begin{equation}
\label{eq:DysonGr_minus_a}
G^r-G^a = g^r - g^a + g^r \Sigma^r G^r - g^a \Sigma^a G^a .
\end{equation}

Using the definitions of the spectral functions $A^G$ and $A^\Sigma$ and the fact that
$\Re e G^r = \Re e G^a = P.V. \int A^G(\omega')/(\omega-\omega')=\Re e G^t$ as in 
Eq.~(\ref{eq:Regt_T_w}) (with similar relations for the self-energy $\Sigma$), we find
\begin{equation}
\label{eq:A_w}
\begin{split}
A^G = A^g & + A^g       \left( \Re e \Sigma . \Re e G - \pi^2 A^\Sigma A^G \right) \\
          & + \Re e (g) . \left( \Re e \Sigma . A^G + A^\Sigma . \Re e G         \right) .
\end{split}
\end{equation}

The time-ordered GF $G^t(\omega)$ can then be obtained from its spectral representation
Eq.~(\ref{eq:gt_T_w}) using the expression of the spectral function $A^G(\omega)$ given
by Eq.~(\ref{eq:A_w}). The first term in the RHS of Eq.~(\ref{eq:A_w}) generates the 
non-interacting time-ordered GF $g^t(\omega)$, while the two other terms regrouped
in $D(\omega)= A^g ( \Re e \Sigma . \Re e G - \pi^2 A^\Sigma A^G )
+ \Re e (g) . ( \Re e \Sigma . A^G + \Re e G . A^\Sigma )$ give a real part
$P.V.  \int D(\omega')/(\omega-\omega')$ and an imaginary part
$\pi [2 f^{\rm eq}(\omega)-1] D(\omega)$.
The proper result for the modified Dyson equation of $G^t(\omega)$ can be obtained
after lengthy algebraic derivations. Instead we now show another just as valid 
shorter route to end up with the correct result.

We consider the different contributions of a conventional Dyson
$g^t + g^t \Sigma^t G^t$. From the spectral representation, we have already generated
the term in $g^t$ from $A^g$, so we are left with the $g^t \Sigma^t G^t$.

We first concentrate on the product $\Sigma^t G^t$:
\begin{equation}
\label{eq:Sigmat_Gt}
\begin{split}
& \Sigma^t G^t  \\
& = \left(\Re e \Sigma + {\rm i} \pi ( 2f^{\rm eq} - 1 )  A^\Sigma \right) 
               \left(\Re e G      + {\rm i} \pi ( 2f^{\rm eq} - 1 )  A^G      \right) \\
& = \left( \Re e \Sigma . \Re e G - \pi^2 A^\Sigma A^G - (2\pi)^2 f^{\rm eq}(f^{\rm eq}-1) A^\Sigma A^G\right) \\
& \qquad + {\rm i} \pi ( 2f^{\rm eq} - 1 ) \left( \Re e \Sigma . A^G + A^\Sigma . \Re e G         \right) \\
& = \left( \Re e \Sigma . \Re e G - \pi^2 A^\Sigma A^G + \Sigma^< G^> \right) \\
& \qquad + {\rm i} \pi ( 2f^{\rm eq} - 1 ) \left( \Re e \Sigma . A^G + A^\Sigma . \Re e G        \right) ,
\end{split}
\end{equation}
where we have introduced, in the last equality, the definition of the product $\Sigma^< G^>$ given
in Eq.~(\ref{eq:SigmaLessGgtr}). We can see that the correction term in $\Sigma^< G^>$ appears
already at this level.

Hence one can see that
\begin{equation}
\label{eq:Sigmat_1plusSigmatGt}
\begin{split}
& 1 + \Sigma^t G^t - \Sigma^< G^> \\
& = \left( 1 + \Re e \Sigma \Re e G - \pi^2 A^\Sigma A^G \right)
 + {\rm i} ( 2f^{\rm eq} - 1 ) \pi  \left( \Re e \Sigma . A^G + A^\Sigma  \Re e G         \right) \ .
\end{split}
\end{equation}

To finish our derivation, we first consider the imaginary part of $g^t( 1 + \Sigma^t G^t - \Sigma^< G^>)$. 
By using the definition for $g^t$ from Eqs.~(\ref{eq:Regt_T_w}) and (\ref{eq:Imgt_T_w})
$g^t = \Re e(g) + i ( 2f^{\rm eq} - 1 ) \pi A^g$, we obtain:
\begin{equation}
\label{eq:ImgtSEtGt}
\begin{split}
& \Im m \left[ g^t( 1 + \Sigma^t G^t - \Sigma^< G^>) \right] \\
& = \Im m (g^t)\ \Re e (1 + \Sigma^t G^t - \Sigma^< G^>) 
+ \Re e(g^t)\ \Im m (1 + \Sigma^t G^t - \Sigma^< G^>) \\
& = ( 2f^{\rm eq} - 1 ) \pi A^g \left( 1 + \Re e \Sigma \Re e G - \pi^2 A^\Sigma A^G \right)
+ \Re e(g) ( 2f^{\rm eq} - 1 ) \pi  \left( \Re e \Sigma . A^G + A^\Sigma  \Re e G   \right) \\
& =( 2f^{\rm eq} - 1 )  \pi A^G \\
& = \Im m G^t \ .
\end{split}
\end{equation}
For the real part of $g^t( 1 + \Sigma^t G^t - \Sigma^< G^>)$, we find that
\begin{equation}
\label{eq:RegtSEtGt}
\begin{split}
& \Re e \left[ g^t( 1 + \Sigma^t G^t - \Sigma^< G^>) \right] \\
& = \Re e (g^t)\ \Re e (1 + \Sigma^t G^t - \Sigma^< G^>) 
 - \Im m (g^t)\ \Im m (1 + \Sigma^t G^t - \Sigma^< G^>) \\
& = \Re e (g) \left( 1 + \Re e \Sigma \Re e G - \pi^2 A^\Sigma A^G \right)
- ( 2f^{\rm eq} - 1 ) \pi  A^g ( 2f^{\rm eq} - 1 ) \pi \left( \Re e \Sigma . A^G + A^\Sigma  \Re e G   \right) \\
& = \Re e (g) \left( 1 + \Re e \Sigma \Re e G - \pi^2 A^\Sigma A^G \right)
 - \pi^2 A^g \left( \Re e \Sigma . A^G + A^\Sigma  \Re e G   \right) \\
& - 4 f^{\rm eq} ( f^{\rm eq} - 1 )\pi  A^g \pi \left( \Re e \Sigma . A^G + A^\Sigma  \Re e G   \right) \ .
\end{split}
\end{equation}
We can use the properties of the Hilbert transform \cite{Bedrosian:1963,Xu:2006,Poularikas:1999} , 
i.e. $\mathcal{H}[\mathcal{H}[f]\ ]=-f$ and
the Bedrosian indentity $\mathcal{H}[fg]]=f\mathcal{H}[g]=g\mathcal{H}[f]$, to show that:
\begin{equation}
\label{eq:RegtSEtGt}
\begin{split}
& \Re e G^t = \mathcal{H}[\pi A^G] \\
& = \mathcal{H}[\pi A^g \left( 1 + \Re e \Sigma \Re e G - \pi^2 A^\Sigma A^G \right)]
+ \mathcal{H}[\pi \Re e(g) \left( \Re e \Sigma . A^G + A^\Sigma  \Re e G   \right) ] \\
& = \mathcal{H}[\pi A^g] \left( 1 + \Re e \Sigma \Re e G - \pi^2 A^\Sigma A^G \right)
+ \mathcal{H}[\pi \Re e(g)] \left( \Re e \Sigma . A^G + A^\Sigma  \Re e G   \right) ] \\
& = \Re e(g) \left( 1 + \Re e \Sigma \Re e G - \pi^2 A^\Sigma A^G \right)
- \pi^2 A^g \left( \Re e \Sigma . A^G + A^\Sigma  \Re e G   \right) ] \ ,
\end{split}
\end{equation}
since $\Re e(g) = \mathcal{H}[\pi A^g]$, and $\mathcal{H}[\pi \Re e(g)] = \pi^2 \mathcal{H}[\mathcal{H}[A^g]\ ]$.

Hence Eq.~(\ref{eq:RegtSEtGt}) becomes
\begin{equation}
\label{eq:Re_Gt}
\begin{split}
\Re e G^t = \Re e \left[ g^t( 1 + \Sigma^t G^t - \Sigma^< G^>) \right] 
+ 4 f^{\rm eq} ( f^{\rm eq} - 1 )\pi  A^g \pi \left( \Re e \Sigma . A^G + A^\Sigma  \Re e G   \right) \ ,
\end{split}
\end{equation}
When considering that, by definition at equilibrium, we have:
\begin{equation}
\label{eq:somedef}
\begin{split}
& g^< = - f^{\rm eq} (g^r - g^a) = {\rm i} f^{\rm eq} 2 \pi A^g \\
& (\Sigma G)^> = -( f^{\rm eq} - 1 ) \left( (\Sigma G)^r - (\Sigma G)^a \right) \ ,
\end{split}
\end{equation}
and $ (\Sigma G)^r - (\Sigma G)^a = \Sigma^r G^r - \Sigma^a G^a 
= -2 {\rm i} (\pi A^\Sigma  \Re e G  +  \Re e \Sigma . \pi A^G)$,
we can see that, as expect, the last term in Eq.~(\ref{eq:Re_Gt}) is just equal to $g^<(\Sigma G)^>$.

Hence we recover once more the modified Dyson equation for the time-ordered GF:
\begin{equation}
G^t(\omega) = g^t(\omega) \left( 1+ \Sigma^t(\omega) G^t(\omega) - \Sigma^<(\omega) G^>(\omega) \right)
- g^<(\omega)(\Sigma(\omega) G(\omega))^>
\end{equation}
in a indirect way from the spectral functions.

We now briefly comment on the imaginary-time temperature Green's functions.
The definition of the temperature GF is, in form, similar to a time-ordered
Green's function:
\begin{equation}
\label{eq:G_T}
\mathcal{G}(\mathbf{x}\tau,\mathbf{x}'\tau')
= - {\rm Tr} \left[\rho_G \mathcal{T_\tau} \Psi(\mathbf{x}\tau) \Psi^\dag(\mathbf{x}'\tau')\right] ,
\end{equation}
where one uses imaginary times $\tau={\rm i}t$ instead of real time $t$, and the $\mathcal{T_\tau}$ orders
the operators according to their value of $\tau$, with the smallest at the right \cite{Abrikosov:1963,Fetter:1971}.

The temperature Green's function $\mathcal{G}(1,2)$ follows a conventional Dyson-like equation:
\begin{equation}
\label{eq:DysonG_T}
\mathcal{G}(1,2) = \mathcal{G}_0(1,2) 
		 + \int {\rm d}(34)\ \mathcal{G}_0(1,3) \Sigma(3,4) \mathcal{G}(4,2) ,
\end{equation}
where the time integrations run over $\tau_i$ from 0 to $\beta$ (and $\mathcal{G}_0$
is the temperature Green's function for the non-interacting system).

At equilibrium, the quantities depend only on the ``time'' difference $\tau_1-\tau_2$
and $\mathcal{G}(\mathbf{x},\mathbf{x}',\tau-\tau')$ is periodic over the range $2\beta$
(or periodic (antiperiodic) over the range $[0,\beta]$ for boson (fermion)). Therefore
$\mathcal{G}$ can be expanded in a Fourier series with coefficients 
$\mathcal{G}(\mathbf{x},\mathbf{x}',\omega_n)$ given in terms of the Matsubara frequencies
$\omega_n$ being integer even (odd) numbers of $\pi/\beta$
for boson (fermion). Each Fourier component obeys also a Dyson equation:
\begin{equation}
\label{eq:DysonG_T_iomega}
\mathcal{G}(\omega_n) = \mathcal{G}_0(\omega_n) + \mathcal{G}_0(\omega_n) \Sigma(\omega_n) \mathcal{G}(\omega_n)
\end{equation}
(in compact notation).

There is a well defined mathematical connection between the temperature GF $\mathcal{G}(\omega_n)$ 
and the real-time GF, this is obtained from the analytical continuation from the complex
frequency ${\rm i}\omega_n$ on the real frequency $\omega$ (and changing, when needed, the sum over the 
Matsubara frequencies onto integrals $1/\beta \sum_{\omega_n}\dots \rightarrow \int {\rm d}\omega/2\pi \dots$).

However, even if $\mathcal{G}$ is an imaginary-time-ordered GF, it does not transform into the
real-time-ordered GF.
The analytical continuation ${\rm i}\omega_n \rightarrow \omega + {\rm i}\eta$ transforms the temperature
GF into the retarded real-time GF (in energy representation)\cite{Abrikosov:1963,Fetter:1971} : 
$\mathcal{G}(\omega_n) \rightarrow G^r(\omega)$. Similarly we have
$\mathcal{G}(\omega_n) \rightarrow G^a(\omega)$ for ${\rm i}\omega_n \rightarrow \omega - {\rm i}\eta$.
This is also consistent with the fact the $G^{r,a}$ obeys a Dyson equation, as much as $\mathcal{G}$
does too.

In order to get the time-ordered GF $G^t(\omega)$ for the temperature GF $\mathcal{G}(\omega_n)$, one
has to adopt the following procedure \cite{Fetter:1971}: from $\mathcal{G}(\omega_n)$ and the 
analytical continuation, one obtains $G^r(\omega)$ and $G^a(\omega)$  from which one extracts the
spectral function $A^G(\omega)$ which is then used to built the time-ordered GF $G^t(\omega)$
from the spectral representation Eq.~(\ref{eq:gt_T_w}).
Following this procedure, we have shown above that one can obtain the modified Dyson equation for 
$G^t(\omega)$.

As another example, we provide in \ref{app:Hubbardmodel} similar calculations for a specific system,
i.e. a finite size system at finite temperature, described by the Hubbard Hamiltonian.

\section{Conclusion}
\label{sec:ccl}

We have derived the dynamical equation for the time-ordered GF $G^t(t,t')$ in the most general case of
non-equilibrium and finite temperature.
In the equilibrium case, we have shown that the corresponding time-ordered GF $G^t(\omega)$ obeys
a Dyson equation only in the limit of zero temperature. At finite temperature, the Dyson equation
for $G^t(\omega)$ needs to be modified by a correction term. We have shown that the correction term
correspond exactly to the corrections introduced {\it ad hoc} by Hedin and Lundqvist in their
treatment of the real-time GF at finite temperature. We have also shown that these corrections
can be obtained from the Matsubara temperature GF following the appropriate protocol for the analytical
continuation from the imaginary frequency onto the real frequency and the spectral representation
of the time-ordered GF. Table \ref{table:recap} summarizes the different GFs formalism and the 
corresponding dynamical equations for the different GFs.

Our work leads to two important conclusions. First, the Keldysh time-loop contour formalism to the
GFs, though developed to deal with non-equilibrium conditions, is also extremely appropriate to
the study of the equilibrium properties of the system. It leads, in a more natural and straightforward 
way, to the correct dynamical equations of the GFs. 
Second, any formalisms based on the resolution of the Dyson equation for the time-ordered GF,
as those used for example in electronic structure and excitations calculations,
are valid only at equilibrium and in the zero temperature limit. 
These formalisms include GF calculations
ranging from the $GW$ approximation \cite{Hedin:1969,Onida:2002,Ness:2011b} 
to more recent work on self-consistent GF equations involving four-point propagators \cite{Starke:2012}.
Beyond the zero temperature, correction terms in the Dyson equation for $G^t(\omega)$ must be taken 
into account, or one needs to work with the Matsubara formalism for $\mathcal{G}(i\omega_n)$.
Such temperature effects were also found for the $T$-matrix approximation in a real-time
formalism in Ref.~\cite{Canivell:1978}.


  \begin{table}
    \centering
    \newcolumntype{C}{>{\centering\arraybackslash}m{0.15\textwidth}}
 \newcolumntype{P}{>{\arraybackslash}p{0.29\textwidth}}
    \begin{tabular*}{\textwidth}{C|P|P|P }
      \hline\noalign{\medskip}
       & Non-equilibrium \newline all $T$   & $T\ne 0$ \newline at equilibrium & $T=0$ \newline at equilibrium \\
\noalign{\medskip}\hline\noalign{\smallskip} \\
$G(\tau, \tau^\prime)$ & Dyson eqn  on $C_K:$
      \newline 
$G = g + g \Sigma G$  &  &  \\\noalign{\medskip}
      $G^r, G^a$ \newline in $(t,t')$ or $(\omega)$ & Dyson eqn: \newline
$G^{r,a} = g^{r,a} + g^{r,a} \Sigma^{r,a} G^{r,a}$ & Dyson eqn: \newline
$G^{r,a} = g^{r,a} + g^{r,a} \Sigma^{r,a} G^{r,a}$ & Dyson eqn: \newline
$G^{r,a} =$ \newline $g^{r,a} + g^{r,a} \Sigma^{r,a} G^{r,a}$  \\ \noalign{\medskip} 
      $G^\lessgtr$ \newline in $(t,t')$ or $(\omega)$  & QKE: \newline
$G^\lessgtr = G^r \Sigma^\lessgtr G^a$ \newline $+(1+G^r \Sigma^r) g^\lessgtr (1+\Sigma^a G^a)$ & QKE: \newline
$G^\lessgtr(\omega)=$ \newline $-f^{\lessgtr,{\rm eq}}(\omega) (G^r-G^a)(\omega)$
& QKE: \newline 
$G^\lessgtr=$ \newline $-f^{\lessgtr,{\rm eq}} (G^r-G^a)$  \\ \noalign{\medskip} 
      $G^t$ \newline in $(t,t')$ or $(\omega)$ & Modified Dyson eqn: \newline
$G^t=g^t(1+\Sigma^t G^t - \Sigma^< G^>)$ \newline + boundary condition & Modified Dyson eqn: \newline
$G^t=g^t(1+\Sigma^t G^t - \Sigma^< G^>)$ \newline $+ g^<(\Sigma G)^>$ & Dyson eqn: \newline
$G^t=g^t + g^t \Sigma^t G^t$ \\ \noalign{\medskip} 
$\mathcal{G}$ \newline in $(i\tau,i\tau')$ or $(i\omega_n)$ & & Dyson eqn: \newline $\mathcal{G}
 = \mathcal{G}_0 + \mathcal{G}_0 \Sigma \mathcal{G}$ & \\ \noalign{\medskip} 
      \lasthline
    \end{tabular*}
    \caption{Summary of the different Green's function formalisms in
      different situations (equilibrium/non-equilibrium, finite/zero
      temperature). The rows show the relevant Green's functions and
      the different dynamical equations (Dyson, modified Dyson, or Quantum kinetic
equation QKE) for each situation. Compact notation is used, involving integration
of space and time (unless the energy $\omega$ representation is used) coordinates.}
\label{table:recap}
  \end{table}


\ack
We thank Martin Stankovksi for pointing us to the early work of Hedin and
Lundqvist on real-time GF at finite-temperature;  Falk Tandetzky, Sangeeta
Sharma and Pina Romaniello for careful reading of the paper and valuable
comments; Arno Schindlmayr for useful comments on the Matsubara formalism; 
and Rex Godby for discussions.

\appendix

\section{Rules for continuation on time-loop contour $C_K$}
\label{app:analyticalcontinuation}

For the following products $P_{(i)}(\tau,\tau')$ on the time-loop contour $C_K$,
\begin{equation}
\label{eq:app_acontinuation_CK}
\begin{split}
P_{(2)}(\tau,\tau')	& =  \int_{C_K} {\rm d}\bar\tau\ A(\tau,\bar\tau) B(\bar\tau,\tau')   \\
P_{(3)}	& =  \int_{C_K} A B C  \\
P_{(n)}	& =  \int_{C_K} A_1 A_2 ... A_n,  \\ 		\nonumber
\end{split}
\end{equation}
we have the following rules for the different components  $P_{(i)}^x(t,t')$ on the real-time axis:
$(x=r,a,>,<)$
\begin{equation}
\label{eq:app_acontinuation_t}
\begin{split}
P_{(2)}^\gtrless(t,t')	& =  \int {\rm d}\bar t\ A^r(t,\bar t) B^\gtrless(\bar t,t') + A^\gtrless(t,\bar t) B^a(\bar t,t')   \\ 
P_{(3)}^\gtrless	& =  \int_t A^\gtrless B^a C^a + A^r B^\gtrless C^a + A^r B^r C^\gtrless  \\
P_{(n)}^r	& =  \int_t A_1^r A_2^r ... A_n^r  \ , \hspace*{5mm}
P_{(n)}^a	 =  \int_t A_1^a A_2^a ... A_n^a.  	\nonumber
\end{split}
\end{equation}

\section{Relationship between the different Green's functions and self-energies}
\label{app:GFSEdefinitions}

The relations between the different components of the Green's
functions and self-energies on the Keldysh time-loop contour are given
as usual, with $X^{\eta_1 \eta_2}(12)\equiv G^{\eta_1 \eta_2}(12), \Sigma^{\eta_1 \eta_2}(12)$ 
or $g^{\eta_1 \eta_2}(12)$.
\begin{equation}
\begin{split}
X^r = X^{t}-X^{<} &= X^{>}-X^{\tilde t} \\
X^a = X^{t}-X^{>} &= X^{<}-X^{\tilde t} \\
X^{t}+X^{\tilde t} &= X^{<}+X^{>}     \\ 
X^{>}-X^{<} &= X^r-X^a
\end{split}
\label{eq:app_gendef}
\end{equation}

The usual lesser and greater projections are defined respectively as
$X^< \equiv X^{+-}$ and $X^> \equiv X^{-+}$, and the usual
time-ordered (anti-time-ordered) as $X^t=X^{++}$ ($X^{\tilde t}=X^{--}$).

\section{Example for a finite system at equilibrium}
\label{app:Hubbardmodel}

In this section, we consider in finite system, with interaction, for which 
analytical expressions
are derivable for the different GFs and we check the validity of Eq.~(\ref{eq:app_Gt_w})
for such a system at equilibrium.

We work with the single-site Hubbard Hamiltonian given by:
\begin{equation}
H = \sum_{\sigma=\uparrow,\downarrow} \varepsilon_{0\sigma} n_\sigma + U n_\uparrow n_\downarrow
\label{eq:HHubbard}
\end{equation}
where $n_\sigma$ is the occupation number operator of the spin $\sigma$.
In the absence of magnetic field, the system is spin degenerate when
$\varepsilon_{0\sigma}=\varepsilon_0$. We consider this case in the following.
The system has 4 different states upon the level occupation $n_\sigma=0,
n_\uparrow=1$ and $n_\downarrow=0$, $n_\uparrow=0$ and $n_\downarrow=1$, 
$n_\uparrow=1$ and $n_\downarrow=1$ with respective energy $0, \varepsilon_0,
\varepsilon_0, 2\varepsilon_0+U$.
One can calculate the advanced and retarded GFs at equilibrium (and finite temperature,
for the finite size system ``connected'' to a thermal bath at temperature $T$)
following the usual prescription \cite{Abrikosov:1963,Fetter:1971}.
Within the Lehmann representation, we find

\begin{equation}
G^{r,a}(\omega) = \frac{1}{Z} \left\{
\frac{ 1+{\rm e}^{-\beta \varepsilon_0}}{\omega-\varepsilon_0 \pm {\rm i}\eta}
+
\frac{{\rm e}^{-\beta \varepsilon_0}+{\rm e}^{-\beta (2\varepsilon_0+U)}}{\omega-\varepsilon_0 - U \pm {\rm i}\eta}
\right\} \ ,
\label{eq:GraHubbard}
\end{equation}
with the partition function $Z=1+2{\rm e}^{-\beta \varepsilon_0}+{\rm e}^{-\beta (2\varepsilon_0+U)}$,
$\beta=1/kT$ and $\eta\rightarrow 0^+$.

We define the thermal occupancy $f$ as follows
\begin{equation}
f   = \frac{{\rm e}^{-\beta \varepsilon_0}+{\rm e}^{-\beta (2\varepsilon_0+U)}}{Z} \ 
{\rm and} \ \ 1-f = \frac{ 1+{\rm e}^{-\beta \varepsilon_0}}{Z} \ .
\label{eq:distrib_f}
\end{equation}

In the non-interacting case ($U=0$), one recovers the usual GF expressions
for a single electronic level, $G^{r,a}=1/(\omega-\varepsilon_0 \pm {\rm i}\eta)$
and $f=f^{\rm eq}=1/(1+{\rm e}^{\beta \varepsilon_0})$ the Fermi distribution
(with the chemical potential $\mu^{\rm eq}=0$ taken as the reference of
energy for all the expression given in this Appendix).

Eq.~(\ref{eq:GraHubbard}) can be rewritten as
\begin{equation}
G^{r,a}(\omega) = 
\frac{ 1 - f}{\omega-\varepsilon_0 \pm {\rm i}\eta}
+
\frac{f}{\omega-\varepsilon_0 - U \pm {\rm i}\eta} \ ,
\label{eq:GraHubbard_withf}
\end{equation}
with the usual relationships between the imaginary and real of $G^{r,a}(\omega)$
based on the properties of the Hilbert transform 
$\mathcal{H}[\pi \delta(\omega-E)]=\mathcal{PV} \frac{1}{(\omega-E)}$ (with $\mathcal{PV}$
standing for the Cauchy principal value).

We now define the self-energy $\Sigma^{r,a}(\omega)$ corresponding to $G^{r,a}(\omega)$
as follows:
\begin{equation}
\frac{ 1 - f}{\omega-\varepsilon_0 \pm {\rm i}\eta}
+
\frac{f}{\omega-\varepsilon_0 - U \pm {\rm i}\eta} = \frac{1}{\omega - \varepsilon_0 - \Sigma^{r,a}(\omega)} \ .
\label{eq:GraHubbard_SE}
\end{equation}
One can separate the real and imaginary parts of both sides of the equality, using
$\Sigma^{r/a}=\Re e \Sigma \pm {\rm i}\Im m \Sigma$:
\begin{equation}
\begin{split}
(1-f)\mathcal{PV} \frac{1}{\omega-\varepsilon_0} + f \mathcal{PV} \frac{1}{\omega-\varepsilon_0-U}
\mp {\rm i}\pi
\left[ (1-f)\delta(\omega-\varepsilon_0) + f \delta(\omega-\varepsilon_0-U) \right] \\
=
\frac{\omega-\varepsilon_0 -\Re e \Sigma \mp {\rm i}\Im m \Sigma}
{(\omega-\varepsilon_0 -\Re e \Sigma)^2+(\Im m \Sigma)^2} \ .
\end{split}
\label{eq:ReIm_SE}
\end{equation}

In the LHS of Eq.~(\ref{eq:ReIm_SE}), the real and imaginary parts are related to each other via
an Hilbert transform as explained above. Hence this must held as well for the RHS of Eq.~(\ref{eq:ReIm_SE}).
Indeed, from Eq.~(\ref{eq:ReIm_SE}), we have
\begin{equation}
\begin{split}
\omega-\varepsilon_0 -\Re e  \Sigma  = 
\left[
(1-f)\mathcal{PV} \frac{1}{\omega-\varepsilon_0} + f \mathcal{PV} \frac{1}{\omega-\varepsilon_0-U}
\right]
\times \mathcal{N}(\omega) \ ,
\end{split}
\label{eq:Re_SE}
\end{equation}
and
\begin{equation}
\begin{split}
\Im m \Sigma = 
\left[ (1-f)\delta(\omega-\varepsilon_0) + f \delta(\omega-\varepsilon_0-U) \right] 
\times \mathcal{N}(\omega) \ ,
\end{split}
\label{eq:Im_SE}
\end{equation}
with $\mathcal{N}^{-1}(\omega)=(\omega-\varepsilon_0 -\Re e \Sigma)^2+(\Im m \Sigma)^2$.
Combining Eqs.(\ref{eq:Re_SE}) and (\ref{eq:Im_SE}), we find that
\begin{equation}
\begin{split}
& \mathcal{N} =  \\
& \left(
\left[
(1-f)\mathcal{PV} \frac{1}{\omega-\varepsilon_0} + f \mathcal{PV} \frac{1}{\omega-\varepsilon_0-U}
\right]^2
+
\left[ (1-f)\delta(\omega-\varepsilon_0) + f \delta(\omega-\varepsilon_0-U) \right]^2
\right) \mathcal{N}^2 \ ,
\end{split}
\label{eq:Numerator_SE}
\end{equation}
hence another definition for $\mathcal{N}(\omega)$.

By using three properties of the Hilbert transform \cite{Bedrosian:1963,Xu:2006,Poularikas:1999},
$\mathcal{H}[\mathcal{H}[g(u)]\ ]=-g(u)$, and the Bedrosian
identity for the Hilbert transform of product functions 
$\mathcal{H}[g(u)h(u)]=g(u) \mathcal{H}[h(u)] = h(u) \mathcal{H}[g(u)]$,
one can show that from Eqs.(\ref{eq:Re_SE}) and (\ref{eq:Im_SE}), we have:
\begin{equation}
\begin{split}
\mathcal{H}[(\omega-\varepsilon_0 -\Re e  \Sigma) \mathcal{N}]
& =  - \mathcal{H}[\Re e  \Sigma] = 
\left((1-f) \mathcal{H}[\mathcal{PV}\frac{1}{\omega-\varepsilon_0}]
+f\mathcal{H} [\mathcal{PV}\frac{1}{\omega-\varepsilon_0-U}]\right) \mathcal{N} \\
& = - \Im m \Sigma \ ,
\end{split}
\label{eq:HT_ImRe_SE}
\end{equation}
hence $\Re e  \Sigma(\omega) = -\mathcal{H}[\Im m \Sigma(\omega)]$ 
as expected, see for example Eq.~(\ref{eq:Regt_T_w}).
The first equality comes from the fact that $\mathcal{H}[(\omega-\varepsilon_0 ) \mathcal{N}]=0$
when $\mathcal{N}^{-1}(\omega=\varepsilon_0)\ne 0$.

Now that we have established the usual relationships between the imaginary and real parts of both
the advanced and retarded GFs and self-energies ($X$) for our finite size model, 
i.e. $X^{r,a}(\omega)=\Re e X(\omega) \mp {\rm i} \Im m X(\omega)$
and $X^r(\omega)-X^a(\omega)=-2 \pi {\rm i} A^X$ (with $A^X(\omega)=\pm \frac{1}{\pi} \Im m X^{r/a}(\omega)$), 
and
$\Re e X(\omega) = \mathcal{PV} \int {\rm d}\omega' \frac{A^X(\omega') }{\omega-\omega'}
= - \mathcal{H}[\Im m X(\omega)]$,
we can go further along and define as usual the lesser and greater components (at equilibrium):
$X^<(\omega)=-f (X^r(\omega)-X^a(\omega)) = 2\pi {\rm i} f A^X(\omega)$,
and $X^>(\omega)=-(f-1) (X^r(\omega)-X^a(\omega)) = 2\pi {\rm i} (f-1) A^X(\omega)$,
as well as the time-ordered components:
$\Re e X^t(\omega) = \Re e X^{r,a}(\omega)$ and $\Im m X^t(\omega)  = \pi (2f-1) A^X(\omega)$.
We recall that for our finite size system at equilibrium, the distribution $f$ is given
by Eq.~(\ref{eq:distrib_f}).

We can now check the validity of Eq.~(\ref{eq:app_Gt_w}) for our model system. For that we follow
the prescriptions given in Section \ref{sec:MatsubaraGF}. We first evaluate the imaginary
part of $g^t+g^t\Sigma^t G^t - g^t \Sigma^< G^>$:

\begin{equation}
\begin{split}
& \Im m \left[ g^t+g^t\Sigma^t G^t - g^t \Sigma^< G^> \right] = \\
& \Im m g^t 
\left(
1+ \Re e \Sigma^t \Re e G^t - \Im m \Sigma^t \Im m G^t - \Sigma^< G^>
\right)
+ \Re e g^t 
\left(\Im m \Sigma^t \Re e G^t + \Re e \Sigma^t \Im m G^t
\right) \ ,
\end{split}
\label{eq:Im_Gt_1}
\end{equation}
since $\Sigma^< G^>$ is real.
Using the different definitions, we find
\begin{equation}
\begin{split}
& \Im m \left[ g^t+g^t\Sigma^t G^t - g^t \Sigma^< G^> \right] = \\
& \Im m g^t 
\left(
1+ \Re e \Sigma \Re e G - \pi^2 (2f-1)^2 A^\Sigma A^G - (2\pi)^2 f(1-f) A^\Sigma A^G
\right) \\
& + \Re e g^t 
\left(\pi (2f-1) A^\Sigma \Re e G + \Re e \Sigma \pi (2f-1) A^G
\right) \ .
\end{split}
\label{eq:Im_Gt_2}
\end{equation}
Using the definition $\Im m g^t = \pi (2f-1) A^g$, we end up with
\begin{equation}
\begin{split}
& \Im m \left[ g^t+g^t\Sigma^t G^t - g^t \Sigma^< G^> \right] = \\
& \pi (2f-1)
\left[ 
A^g
\left(
1+ \Re e \Sigma \Re e G - \pi^2 A^\Sigma A^G \right)
+ \Re e g 
\left( A^\Sigma \Re e G + \Re e \Sigma A^G \right) 
\right] ,
\end{split}
\label{eq:Im_Gt_3}
\end{equation}
where the term in brakets is just the definition of the spectral function $A^G$
given by Eq.~(\ref{eq:A_w}).
Hence we find that
\begin{equation}
\begin{split}
& \Im m \left[ g^t+g^t\Sigma^t G^t - g^t \Sigma^< G^> \right] = \pi (2f-1) A^G = \Im m G^t \ ,
\end{split}
\label{eq:Im_Gt_4}
\end{equation}
as expected.
One can proceed along the same lines to establish the relation for the real parts, using the
different properties of the Hilbert transform mentioned above. 
Therefore, as expected, Eq.~(\ref{eq:app_Gt_w}) is valid for the finite size system, at
equilibrium, described by Hamiltonian Eq.~(\ref{eq:HHubbard}).

\section*{References}

\providecommand{\newblock}{}


\begin{thebibliography}{10}
\expandafter\ifx\csname url\endcsname\relax
  \def\url#1{{\tt #1}}\fi
\expandafter\ifx\csname urlprefix\endcsname\relax\def\urlprefix{URL }\fi
\providecommand{\eprint}[2][]{\url{#2}}

\bibitem{Pines:1961}
Pines D 1961 {\em The Many-Body Problem\/} (New York: W.A. Benjamin)

\bibitem{Kadanoff:1962}
Kadanoff L~P and Baym G 1962 {\em {Q}uantum {S}tatistical {M}echanics\/} (New
  York: W.A. Benjamin)

\bibitem{Abrikosov:1963}
Abrikosov A~A, Gorkov L~P and Dzyaloshinski I~E 1963 {\em {M}ethods of
  {Q}uantum {F}ield {T}heory in {S}tatistical {P}hysics\/} (New York: Dover)

\bibitem{Fetter:1971}
Fetter A~L and Walecka J~D 1971 {\em {Q}uantum {T}heory of {M}any-{P}article
  {S}ystems\/} (New York: McGraw-Hill)

\bibitem{Rammer:1986}
Rammer J and Smith H 1986 {\em Review of Modern Physics\/} {\bf 58} 323

\bibitem{Mahan:1990}
Mahan G~D 1990 {\em {M}any-{P}article {P}hysics\/} (New York: Plenum Press)

\bibitem{Bruus:2004}
Bruus H and Flensberg K 2004 {\em {M}any-body {Q}uantum {T}heory in {C}ondensed
  {M}atter\/} (Oxford: Oxford University Press)

\bibitem{Dickhoff:2008}
Dickhoff W~H and Neck D~V 2008 {\em Many-body theory exposed!\/} (Singapore:
  World Scientific)

\bibitem{Keldysh:1965}
Keldysh L 1965 {\em Sov. Phys. JETP\/} {\bf 20} 1018

\bibitem{Danielewicz:1984}
Danielewicz P 1984 {\em Annals of Physics\/} {\bf 152} 239

\bibitem{Chou:1985}
chao Chou K, bin Su Z, lin Hao B and Yu L 1985 {\em Physics Reports\/} {\bf
  118} 1

\bibitem{Wagner:1991}
Wagner M 1991 {\em Physical Review B\/} {\bf 44} 6104

\bibitem{vanLeeuwen:2006}
van Leeuwen R, Dahlen N~E, Stefanucci G, Almbladh C~O and von Barth U 2006 {\em
  Lecture Notes in Physics\/} {\bf 706} 33

\bibitem{Rammer:2007}
Rammer J 2007 {\em Quantum Field Theory of Non-Equilibrium States\/}
  (Cambridge: Cambridge University Press)

\bibitem{Schwinger:1961}
Schwinger J 1961 {\em J. Math. Phys.\/} {\bf 2} 407

\bibitem{Hedin:1969}
Hedin L and Lundqvist S 1969 {\em Effects of Electron-Electron and
  Electron-Phonon Interactions on the One-Electron States of Solids\/} vol~23
  (New York: Academic Press)

\bibitem{Stefanucci:2004a}
Stefanucci G and Almbladh C O 2004 {\em Phys. Rev. B} {\bf 69} {195318}

\bibitem{Doyon:2006}
Doyon B and Andrei N 2006 {\em Phys. Rev. B} {\bf 73} {245326}

\bibitem{Stefanucci:2007}
Stefanucci G 2007 {\em Phys. Rev. B} {\bf 75} {195115}

\bibitem{footnote2}
{We have also found that when calculating the retarded GF $G^r=G^t-G^<$
from Eq.~(\ref{eq:app_Gt_last_bis}) for $G^t$ and 
Eq.~(\ref{eq:DysonGlessgtr}) for $G^<$, we recover the conventional Dyson equation
for $G^r$ but only thanks to the presence of the term $\Sigma ^< G^>$ in the definition
of $G^t$.}

\bibitem{footnote3}
{One can check that Eq. (8-27a) and Eq. (8-27b) in Ref.~\cite{Kadanoff:1962}
are equivalent to the equation of motion Eq.~(\ref{eq:EOM}) for $G^\lessgtr$, i.e.
$L(1)G^\lessgtr(1,1') = (\Sigma G)^\lessgtr(1,1')$ with the one-point operator
$L(1)=i\partial_{t_1} - \bar{h}_0(1)$ and the relation 
$(XY)^\lessgtr = X^r Y^\lessgtr + X^\lessgtr Y^a$. Then by using the definition of the
time-ordered GF $G^t(1,2) = - {\rm i} \langle \mathcal{T} \Psi(1) \Psi^\dag(2) \rangle$
given in terms of $G^\lessgtr$: $G^t(1,2)=\theta(t_1-t_2) G^>(1,2) + \theta(t_2-t_1) G^<(1,2)$
and calculating directly its equation of motion $L(1)G^t(1,2)$, one can recover the
expression of the modified Dyson equation Eq.~(\ref{eq:app_Gt_last_ter}).}

\bibitem{Lipavski:1986}
Lipavsk\'y P, \v{S}pi\v{c}ka V and Velick\'y B 1986 {\em Physical Review B\/}
  {\bf 34} 6933

\bibitem{Meden:1995}
Meden V, W\"ohler C, Fricke J and Sch\"onhammer K 1995 {\em Physical Review
  B\/} {\bf 52} 5624

\bibitem{Vasko:2005}
Vasko F~T and Raichev O~E 2005 {\em Quantum Kinetic Theory and Applications:
  Electrons, Photons, Phonons\/} (New York: Springer Science+Business Media
  Inc)

\bibitem{Ness:2010}
Ness H, Dash L and Godby R~W 2010 {\em Physical Review B\/} {\bf 82} 085426

\bibitem{Kita:2010}
Kita T 2010 {\em Progress of Theoretical Physics\/} {\bf 123} 581

\bibitem{Bedrosian:1963}
Bedrosian E, 1963 {\em Proc. IEEE} {\bf 51} 868

\bibitem{Xu:2006}
Xu Y and Yan D 2006 {\em Proc. Amer. Math. Soc.} {\bf 134} 2719

\bibitem{Poularikas:1999}
Poularikas A. D. 1999 {\em The Handbook of Formulas and Tables for Signal Processing}
(Boca Raton: CRC Press LLC). Chapter 15 on ``The Hilbert Transform''.

\bibitem{Onida:2002}
Onida G, Reining L and Rubio A 2002 {\em Rev. Mod. Phys.\/} {\bf 74} 601--659

\bibitem{Ness:2011b}
Ness H, Dash L~K, Stankovski M and Godby R~W 2011 {\em Physical Review B\/}
  {\bf 84} 195114

\bibitem{Starke:2012}
Starke R and Kresse G 2012 {\em Physical Review B\/} {\bf 85} 075119

\bibitem{Canivell:1978}
Canivell V, Garrido L, Miguel M~S and Seglar P 1978 {\em Physical Review A\/}
  {\bf 17} 480


\end{thebibliography}
\end{document}